\documentclass[twocolumn,twoside,slac]{revtex4}
\usepackage{graphicx}
\usepackage{fancyhdr}
\begin{document}

\title{Measurement of normalized differential $t\bar{t}$ cross sections in the dilepton channel in pp collisions at center-of-mass energy of 13 TeV}

%

\author{Youn Jung Roh}
\affiliation{Department of Physics, Korea University, Seoul, Korea}
\author{On behalf of the CMS Collaboration}

\begin{abstract}
Measurements of normalized differential cross sections for top quark pair production are performed in the dilepton decay channels in proton-proton collisions at a center-of-mass energy of 13 TeV. The differential cross sections are measured with data corresponding to an integrated luminosity of 2.1 fb$^{-1}$ recorded by the CMS experiment at the LHC. We have measured the cross sections differentially as a function of the kinematic properties of the leptons (electron or muon), jets from bottom quark hadronization, top quarks, and top quark pairs at the particle and parton levels. The $t\bar{t}$ differential cross section measurements are compared to several Monte Carlo generators that implement calculations up to next-to-leading order in perturbative quantum chromodynamics interfaced with parton showering, and also to fixed-order theoretical calculations of top quark pair production beyond next-to-leading order accuracy.
\end{abstract}

\maketitle

\thispagestyle{fancy}


\section{Introduction}

\begin{figure}[htb]
\begin{center}
\scalebox{0.1}
     {
      \includegraphics{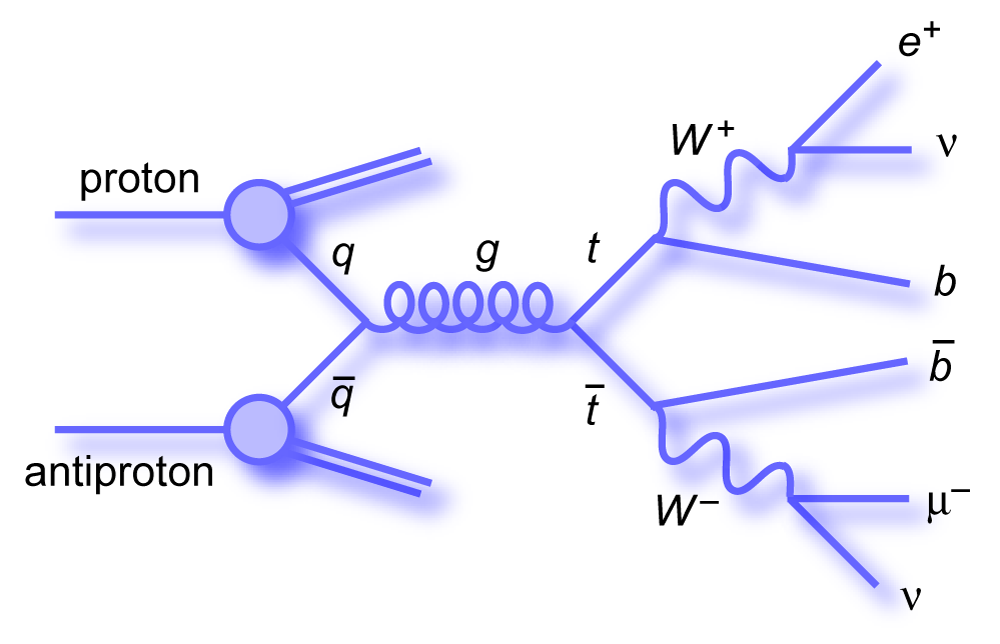}
     }
\end{center}
\label{fig:fey}
\caption{Feynman diagram of top-quark pair production decaying to lepton pairs.}

 \end{figure}

Measurements of top quark pair ($t\bar{t}$) production cross sections as a function of top quark related kinematic observables are crucial for testing perturbative quantum chromodynamics (QCD) calculations and for probing a variety of different properties of the top quark. Moreover, they can reveal hints of new physics phenomena beyond the Standard Model. In this document, recent results of normalized $t\bar{t}$ differential cross sections measured in dilepton (electron or muon) final states (see Figure~\ref{fig:fey}) are presented. The measurements are performed using proton-proton collision data produced at the CERN LHC at a center-of-mass energy of 13 TeV and recorded by the CMS experiment \cite{cms} in 2015. The analyzed data correspond to an integrated luminosity of 2.1 fb$^{-1}$ .

The measurements are performed at the particle and parton levels. The visible particle-level measurements use final-state objects that are experimentally measurable and theoretically well defined to minimize Monte Carlo (MC) modeling dependence and to avoid large extrapolation, so variables are corrected mainly for detector effects. In contrast, the parton-level measurements are derived in the full phase space to compare to predictions of perturbative QCD beyond next-to-leading order (NLO) accuracy.

\section{Data samples}

Double lepton (electron or muon) triggered data are used for this analysis. MC techniques are used to simulate the signal and background processes. The simulation includes $t\bar{t}$+jets, Z/$\gamma^{*}$+jets, W+jets, single top quark production and diboson (WW,ZZ,WZ) processes.
Simulated samples are generated by POWHEG and MG5$\_$aMC$@$NLO and showered with either PYTHIA8 or HERWIG++. 

\section{Signal definition and event selection} 

The measured $t\bar{t}$ differential cross sections are presented at both particle and parton levels as a function of kinematic observables of the top quarks and the ttbar system, defined at generator level. The particle-level top quark is defined at the generator level using the definitions of the final-state objects described as follows:

\begin{itemize}
\item Prompt neutrino : neutrinos not from hadron decays 
\item Dressed lepton : anti-kt algorithm with a distance parameter of 0.1 using electrons, muons and photons not from hadron decays,  \\$p_{T} >$ 20 GeV, $|\eta| <$ 2.4 

\item b quark jet : anti-kt algorithm with a distance parameter of 0.4 using all particles and ghost-B hadrons not including any neutrinos nor particles used in dressed leptons, $p_{T} >$ 30 GeV, \\$|\eta| <$ 2.4

\end{itemize}

A W boson at the particle level is reconstructed by combining a dressed lepton and a prompt neutrino. A pair of particle-level W bosons is chosen among the possible combinations to minimise the scalar sum of invariant mass differences with respect to the W boson mass of 80.4 GeV. Similarly, the top quark at the particle-level is defined by combining a particle-level W boson and a b quark jet, with the minimum invariant mass difference from the correct top quark mass of 172.5 GeV. The visible phase space is defined to have a pair of particle-level top quarks, constructed from prompt neutrinos, dressed leptons, and b jets.

In addition, the parton-level objects are defined before the top quark decays into a bottom quark and a W boson and after QCD radiation. The normalized differential cross sections at the parton level are derived by extrapolating the measurements into the full phase space.

The dilepton decay channels consist of two leptons, at least two jets, and missing transverse energy ($p_{\rm T}^{\rm miss}$) from two neutrinos. Events are selected using dilepton triggers, and additional selections are applied to filter signal event as follows: 

\begin{itemize}
\item \textbf{Electron} : $p_{\rm T} >$ 20 GeV, $ |\eta | <$2.4
\item \textbf{Muon} : $p_{\rm T} >$ 20 GeV, $ |\eta | <$2.4

\item \textbf{Jet} :$p_{\rm T} > $ 30 GeV, $ |\eta | <$2.4

\item Two opposite-charged leptons ($ee/\mu\mu/e\mu$) with invariant mass of the lepton pair $M_{ll} >$ 20 GeV
\item Z mass veto ($| M_{ll} - 91|<$ 15 GeV) for $ee/\mu\mu$  
\item Missing transverse energy  $p_{\rm T}^{\rm miss} > $ 40 GeV for $ee/\mu\mu$ 
\item Requirement of 2 jets and 1 b-tagged jet
\end{itemize}

\section{Top reconstruction and normalized differential cross section}

The top quarks are reconstructed using the four momenta of all final-state objects by an algebraic kinematic reconstruction method.  Constraints such as the balance of $p_{\rm T}$ of the two neutrinos and mass of the W boson and the top quark are imposed.  The $t\bar{t}$ system is reconstructed for 100 different random variations within their simulated resolution functions and varying the W boson mass to consider effects of detector resolution. 

\begin{figure}[htb]

\begin{center}

\vspace{5pt}
\scalebox{0.6}
{

   \includegraphics[width=0.4\textwidth]{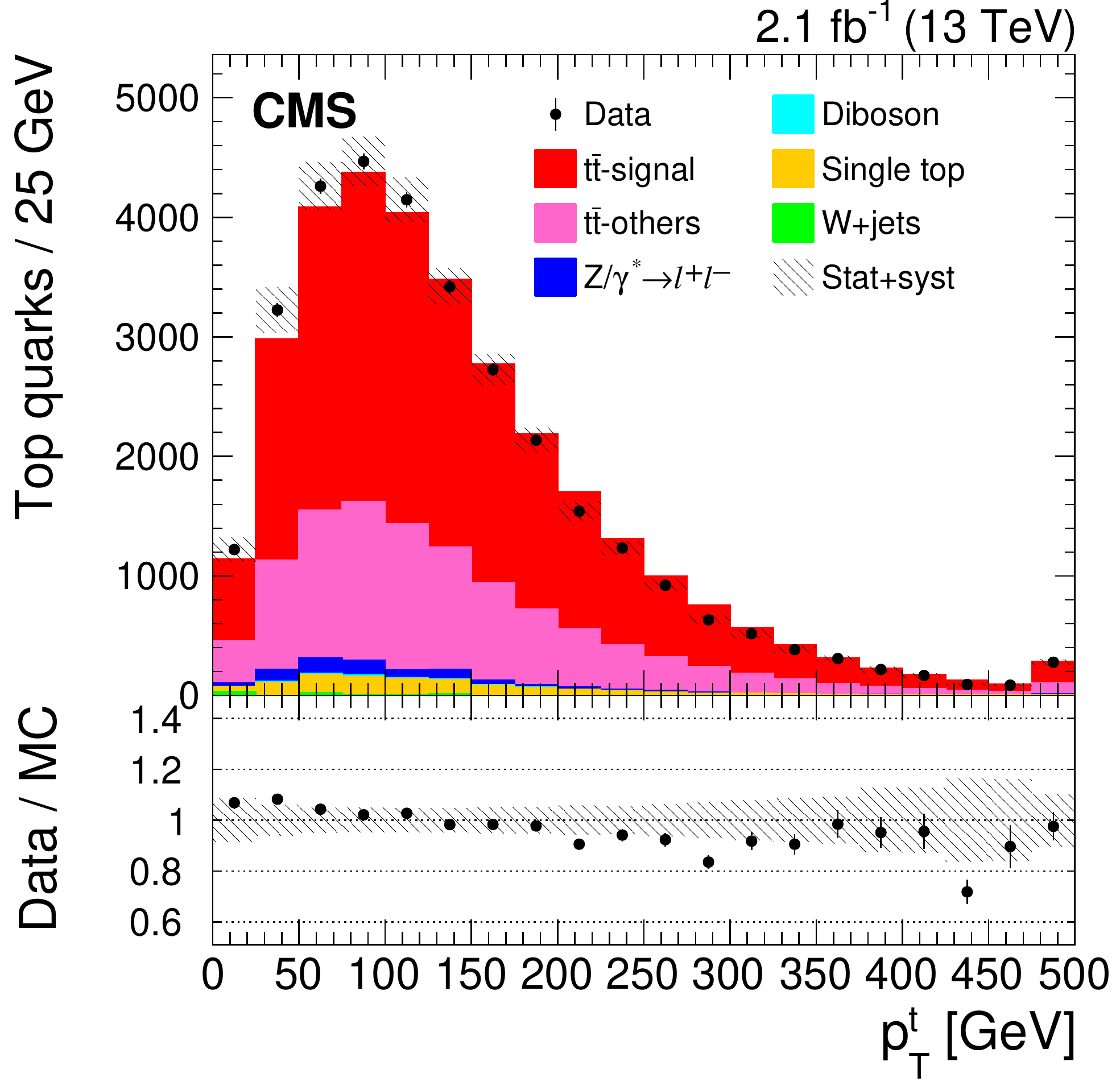}
   \includegraphics[width=0.4\textwidth]{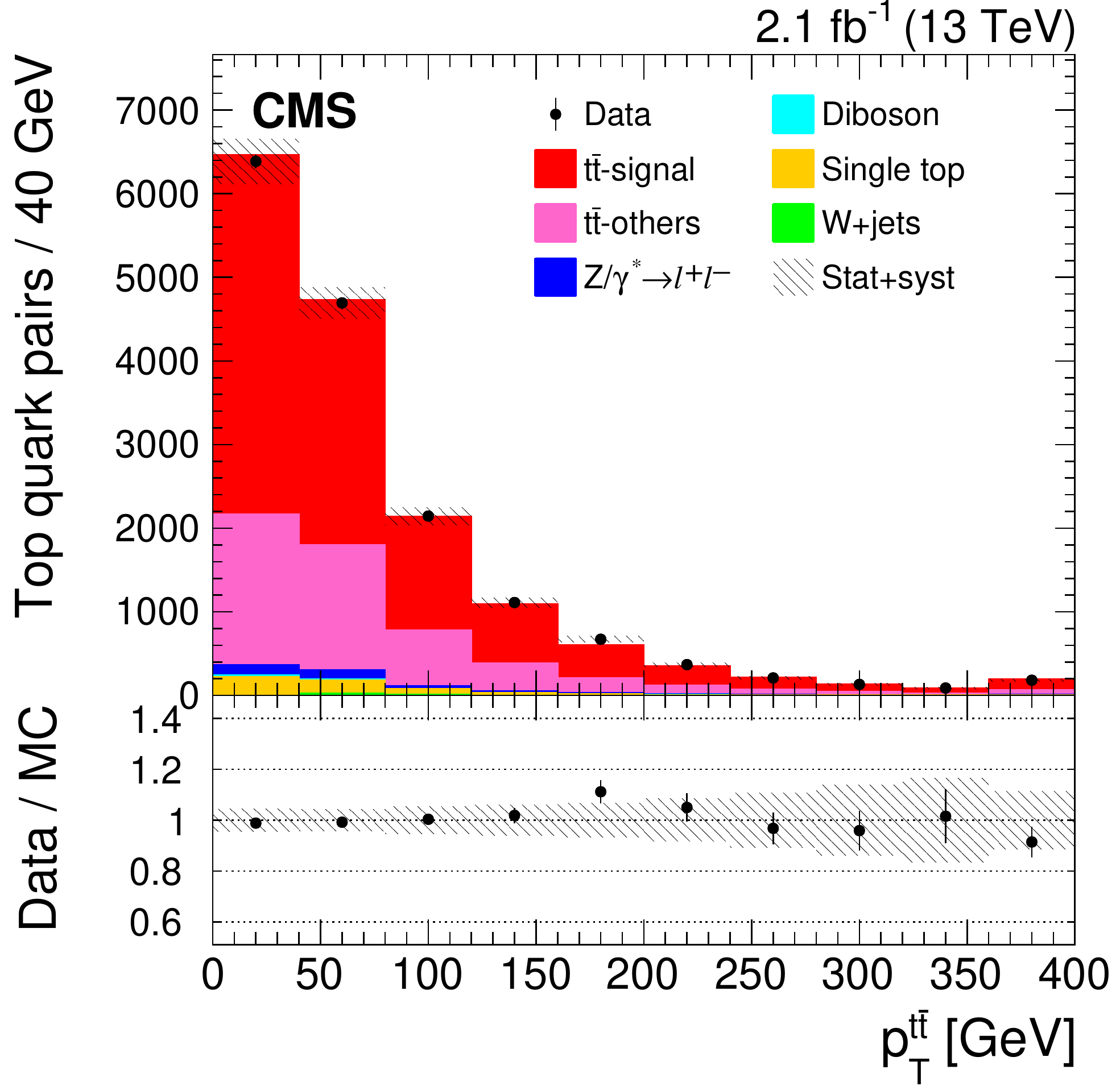}
      
}

\end{center}
\caption{Kinematic distributions after event selection and the kinematic reconstruction 
of the $t \bar{t}$ system. Figure taken from~\cite{paper}}

\label{fig:top_reco}
 \end{figure}

In each trial, the minimum invariant mass of the ${t\bar{t}}$ system are selected, and a weight is calculated using the expected invariant mass distribution of lepton and b jet pairs. The lepton and b jet pairs with the maximum sum of weights are chosen, and the neutrino momentum is determined using the weighted average over the trials. Figure~\ref{fig:top_reco} displays the distribution of the transverse momenta of top quark ($p_{\rm T}^{\rm t}$) and of top quark pair ($p_{\rm T}^{\rm t \bar{t}}$).

The normalized differential $t\bar{t}$ cross sections $ (1/ \sigma)( \rm {d} \sigma / \rm {d} X )$ are measured as a function of several kinematic variables $X$. The corrections for detector efficiencies, acceptances, and migrations are performed using a D'Agostini unfolding method.
$\sigma$ is the total cross section. $X$ represents a variable such as  
top $p_{\rm T}$,  $p_{\rm T}^{\rm t\bar{t}}$, $t\bar{t}$ mass, and  $\Delta \phi_{t\bar{t}}$.

\section{Results}
\begin{figure}[htb]

\begin{center}

\scalebox{0.6}
     {
    \includegraphics[width=0.6\textwidth]{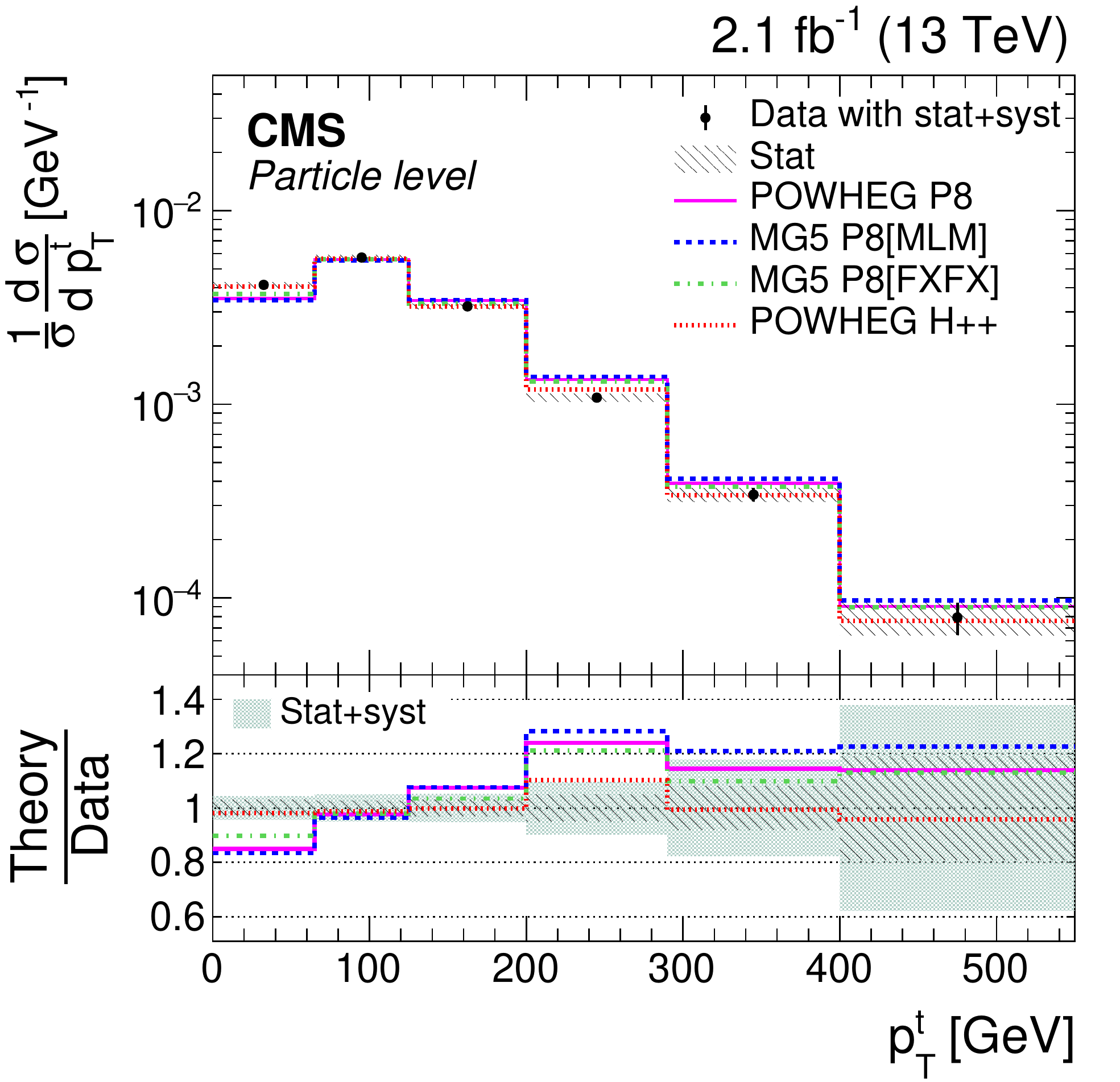}
    \includegraphics[width=0.6\textwidth]{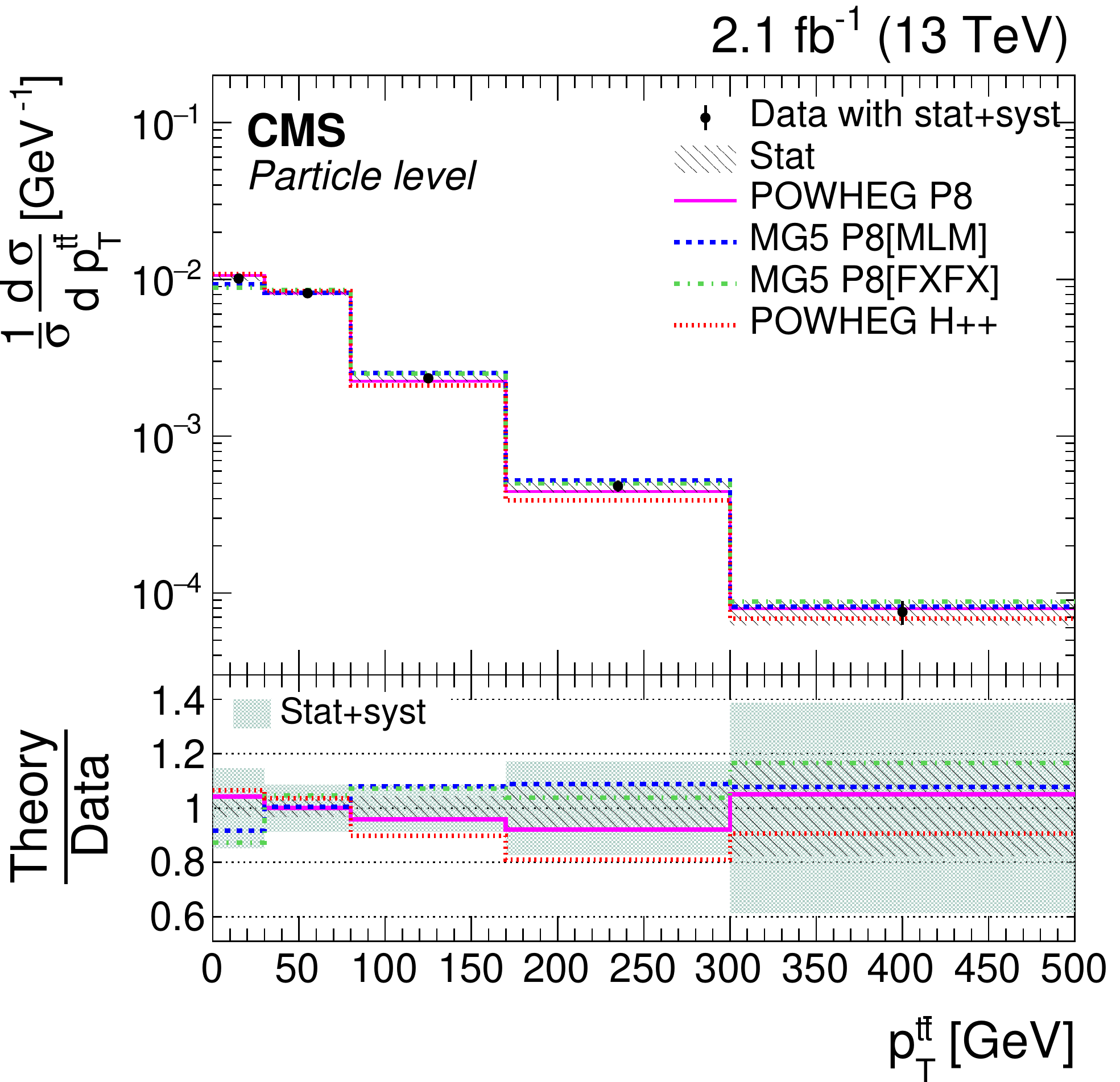}
    
     }

\scalebox{0.6}
{

  \includegraphics[width=0.6\textwidth]{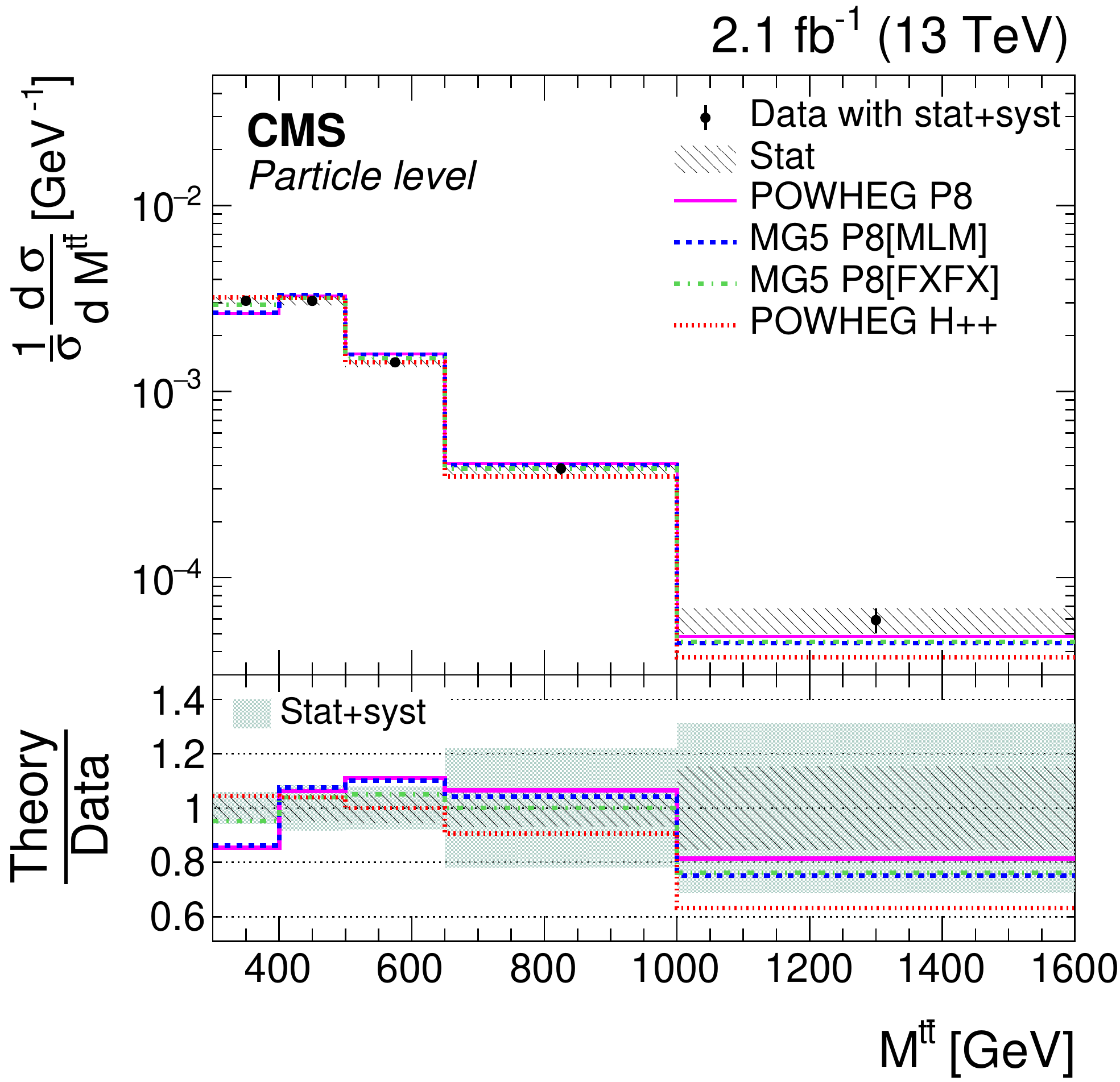}
  \includegraphics[width=0.6\textwidth]{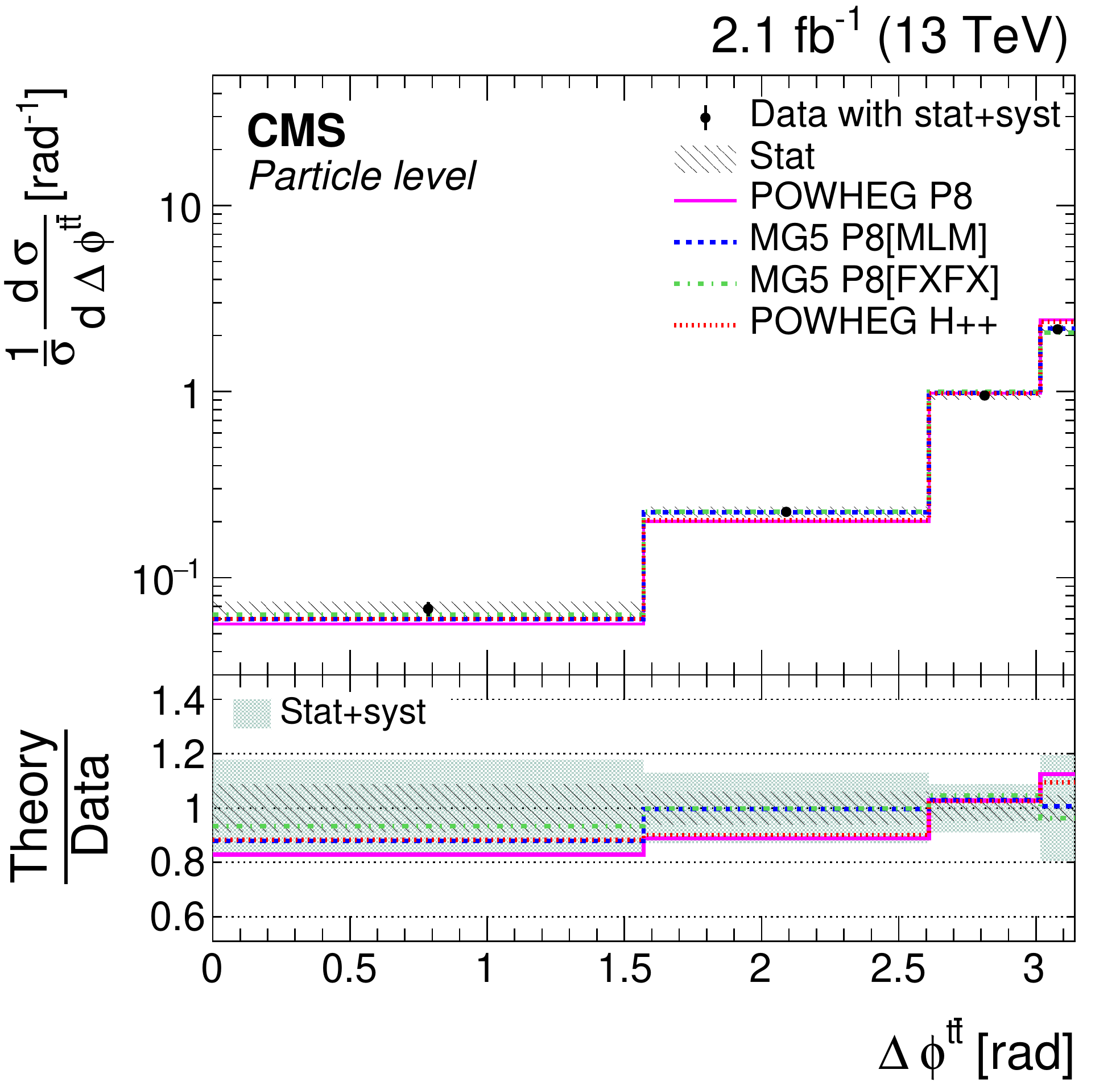}
      
}
\scalebox{0.6}
{

   \includegraphics[width=0.6\textwidth]{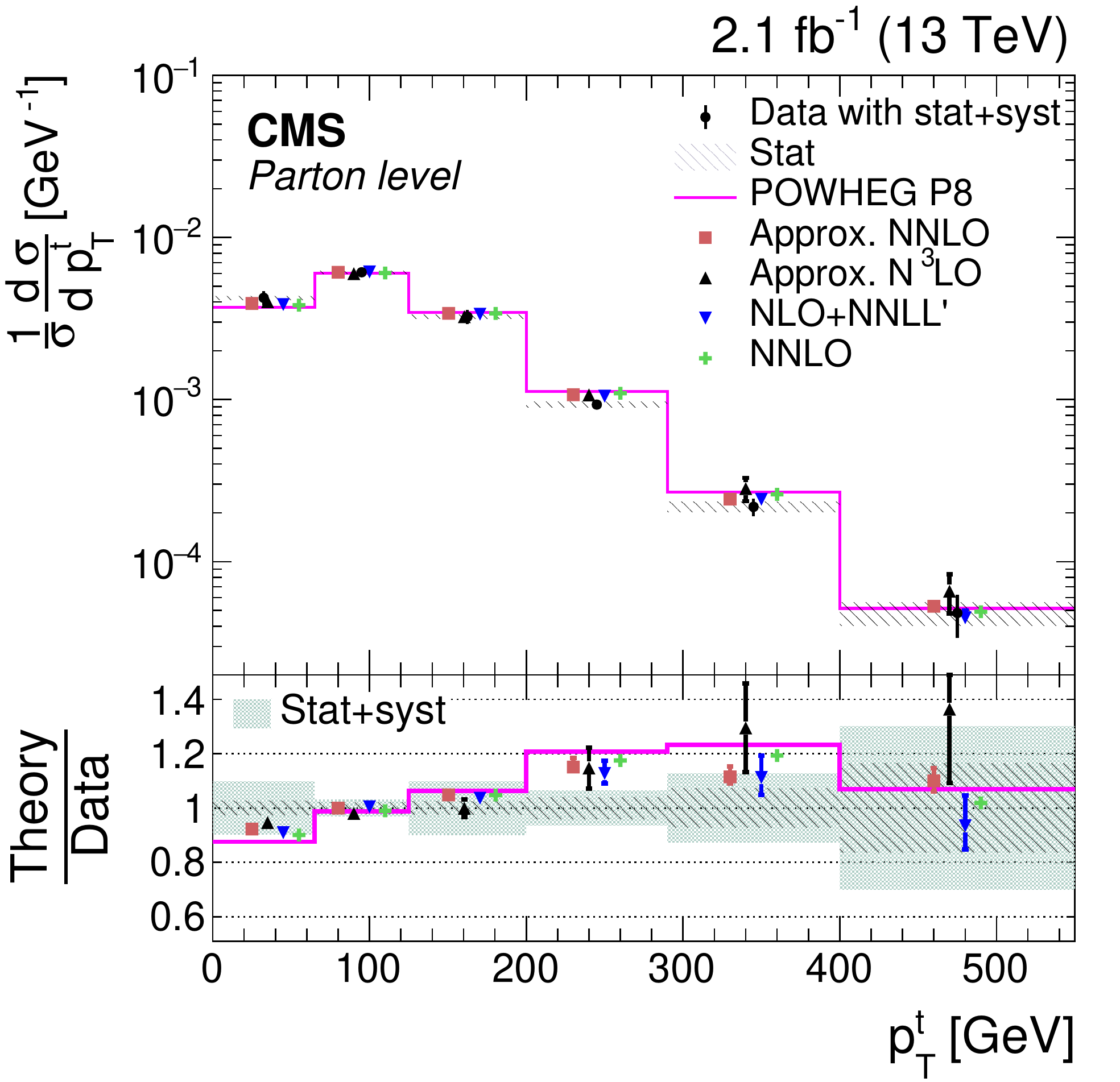}
   \includegraphics[width=0.6\textwidth]{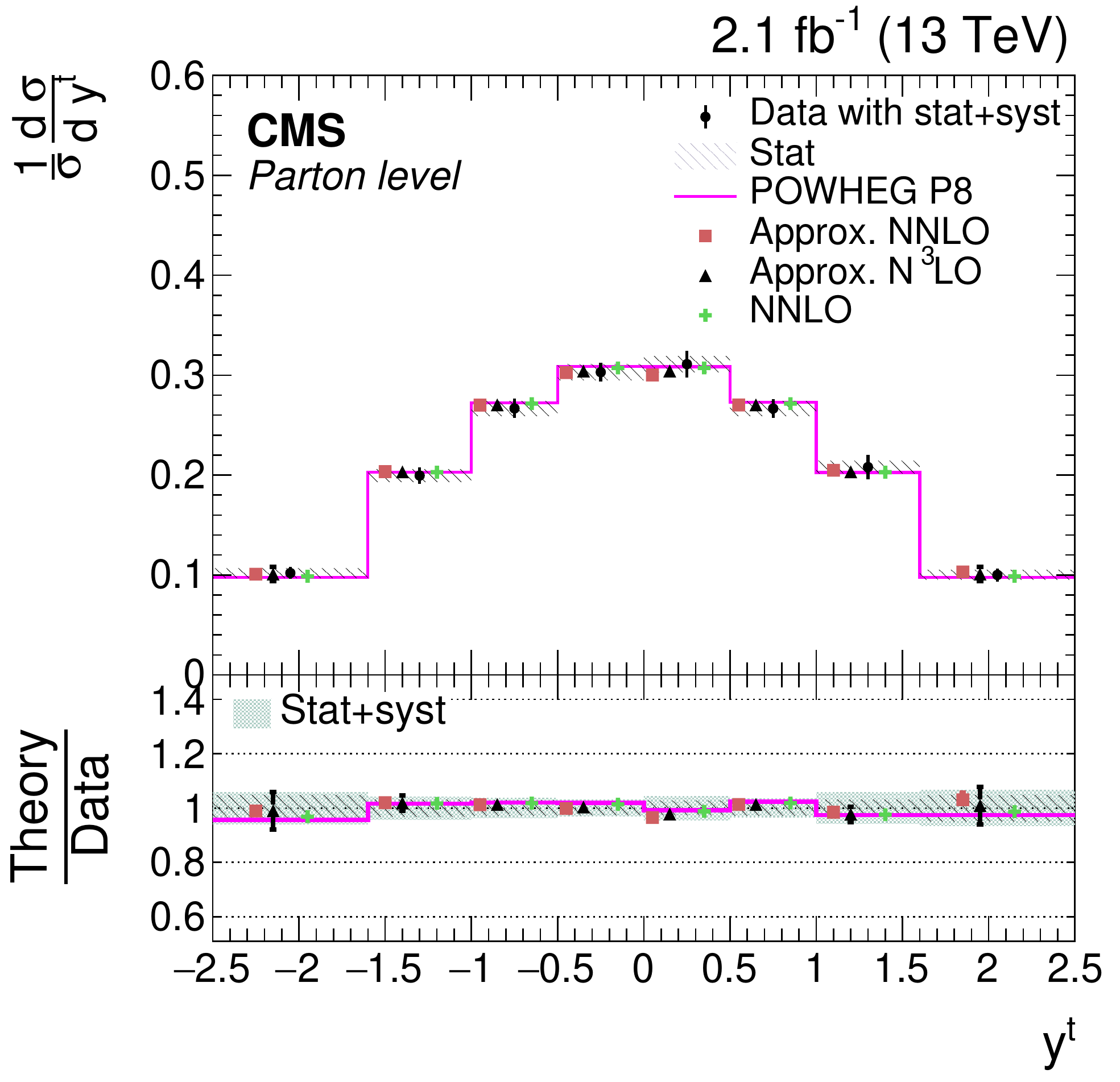}
      
}

\end{center}
\caption{Normalized differential $t\bar{t}$ cross sections as a function of the top $p_{\rm T}$ (upper left), $p_{\rm T}^{\rm t\bar{t}}$ (upper right), and $t\bar{t}$ mass (middle left) and $\Delta \phi_{t\bar{t}}$ (middle right), measured at the particle level in the visible phase space. The measured data are compared to different MC predictions (see text). Normalized differential $t\bar{t}$ cross sections as a function of top $p_{\rm T}$ (bottem left) and $p_{\rm T}^{\rm t\bar{t}}$ (bottom right), measured at the parton level in the full phase space and compared to different perturbative QCD calculations beyond NLO accuracy. The vertical bars on the data points indicate the total (combined statistical and systematic) uncertainties while the hatched band shows the statistical uncertainty. The lower panel gives the ratio of the theoretical predictions to the data. The light-shaded band displays the combined statistical and systematic uncertainties added in quadrature. Figure taken from~\cite{paper}}

\label{fig:result}
 \end{figure}

The normalized differential cross sections at particle level are measured as a function of the top $p_{\rm T}$,  $p_{\rm T}^{\rm t\bar{t}}$, $t\bar{t}$ mass, and  $\Delta \phi_{t\bar{t}}$, shown in Figure~\ref{fig:result} (top and middle rows). Figure~\ref{fig:result} (bottom) presents the normalized differential $t\bar{t}$ cross sections as a function of top $p_{\rm T}$ and $p_{\rm T}^{\rm t\bar{t}}$, measured at the parton level in the full phase space and compared to different perturbative QCD calculations of an approximate next-to-next-to-leading order (NNLO)~\cite{Guzzi}, an approximate next-to-NNLO (N$^{3}$LO)~\cite{Kidonakis}, an improved NLO and next-to-next-to-leading-logarithmic (NLO+NNLL')~\cite{Pecjak}, and  a full NNLO~\cite{Czakon}. 

\section{Conclusions}

Normalized differential cross sections of top quark pair production 
in the dilepton decay channel are measured at the particle level in the visible phase space and the parton level in the full phase space with respect to the top $p_{\rm T}$,  $p_{\rm T}^{\rm t\bar{t}}$, $t\bar{t}$ mass, and  $\Delta \phi_{t\bar{t}}$. The measured differential cross sections are found to be 
in agreement with the standard model predictions, being the top quark $p_{\rm T}$ distribution the only one observed to be in mild tension with the NLO predictions. More details can be found in~\cite{paper}.

\begin{acknowledgments}

This study was carried out with financial support of the National Research Foundation of Korea (NRF), funded by the Ministry of Science $\&$ ICT under contract NRF-2008-00460. It was also supported by the Basic Science Research Program through the National Research Foundation of Korea (NRF) funded by the Ministry of Education (NRF-2016R1A6A3A11933762) and by a Korea University Grant.

\end{acknowledgments}

\end{document}